\documentclass[twoside,twocolumn,9pt]{article}
\usepackage{extsizes}
\usepackage[super,sort&compress,comma]{natbib} \usepackage{xcolor}

\usepackage[version=3]{mhchem}
\usepackage[left=1.5cm, right=1.5cm, top=1.785cm, bottom=2.0cm]{geometry}
\usepackage{balance}
\usepackage{mathptmx}
\usepackage{sectsty}
\usepackage{graphicx} 
\usepackage{lastpage}
\usepackage[format=plain,justification=justified,singlelinecheck=false,font={stretch=1.125,small,sf},labelfont=bf,labelsep=space]{caption}
\usepackage{float}
\usepackage{fancyhdr}
\usepackage{fnpos}
\usepackage[english]{babel}
\addto{\captionsenglish}{%
  
}
\usepackage{array}
\usepackage{droidsans}
\usepackage{charter}
\usepackage[T1]{fontenc}
\usepackage{setspace}
\usepackage[compact]{titlesec}
\usepackage{hyperref}
\usepackage{amsmath,amssymb}

\usepackage{epstopdf}

\definecolor{cream}{RGB}{222,217,201}

\begin{document}

\pagestyle{fancy}
\thispagestyle{plain}
\fancypagestyle{plain}{
\renewcommand{\headrulewidth}{0pt}
}

\makeFNbottom
\makeatletter
\renewcommand\LARGE{\@setfontsize\LARGE{15pt}{17}}
\renewcommand\Large{\@setfontsize\Large{12pt}{14}}
\renewcommand\large{\@setfontsize\large{10pt}{12}}
\renewcommand\footnotesize{\@setfontsize\footnotesize{7pt}{10}}
\makeatother

\renewcommand{\thefootnote}{\fnsymbol{footnote}}
\renewcommand\footnoterule{\vspace*{1pt}%
\color{cream}\hrule width 3.5in height 0.4pt \color{black}\vspace*{5pt}} 
\setcounter{secnumdepth}{5}

\makeatletter 
\renewcommand\@biblabel[1]{#1}            
\renewcommand\@makefntext[1]%
{\noindent\makebox[0pt][r]{\@thefnmark\,}#1}
\makeatother 
\renewcommand{\figurename}{\small{Fig.}~}
\sectionfont{\sffamily\Large}
\subsectionfont{\normalsize}
\subsubsectionfont{\bf}
\setstretch{1.125} 
\setlength{\skip\footins}{0.8cm}
\setlength{\footnotesep}{0.25cm}
\setlength{\jot}{10pt}
\titlespacing*{\section}{0pt}{4pt}{4pt}
\titlespacing*{\subsection}{0pt}{15pt}{1pt}

\fancyfoot{}
\fancyfoot[LO,RE]{\vspace{-7.1pt}\includegraphics[height=9pt]{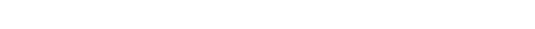}}
\fancyfoot[CO]{\vspace{-7.1pt}\hspace{11.9cm}\includegraphics{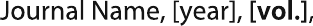}}
\fancyfoot[CE]{\vspace{-7.2pt}\hspace{-13.2cm}\includegraphics{head_foot/RF}}
\fancyfoot[RO]{\footnotesize{\sffamily{1--\pageref{LastPage} ~\textbar  \hspace{2pt}\thepage}}}
\fancyfoot[LE]{\footnotesize{\sffamily{\thepage~\textbar\hspace{4.65cm} 1--\pageref{LastPage}}}}
\fancyhead{}
\renewcommand{\headrulewidth}{0pt} 
\renewcommand{\footrulewidth}{0pt}
\setlength{\arrayrulewidth}{1pt}
\setlength{\columnsep}{6.5mm}
\setlength\bibsep{1pt}

\makeatletter 
\newlength{\figrulesep} 
\setlength{\figrulesep}{0.5\textfloatsep} 

\newcommand{\topfigrule}{\vspace*{-1pt}%
\noindent{\color{cream}\rule[-\figrulesep]{\columnwidth}{1.5pt}} }

\newcommand{\botfigrule}{\vspace*{-2pt}%
\noindent{\color{cream}\rule[\figrulesep]{\columnwidth}{1.5pt}} }

\newcommand{\dblfigrule}{\vspace*{-1pt}%
\noindent{\color{cream}\rule[-\figrulesep]{\textwidth}{1.5pt}} }

\makeatother

\twocolumn[
  \begin{@twocolumnfalse}
{\includegraphics[height=30pt]{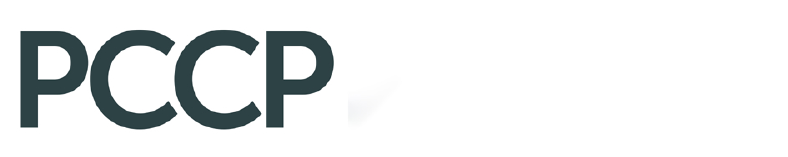}\hfill\raisebox{0pt}[0pt][0pt]{\includegraphics[height=55pt]{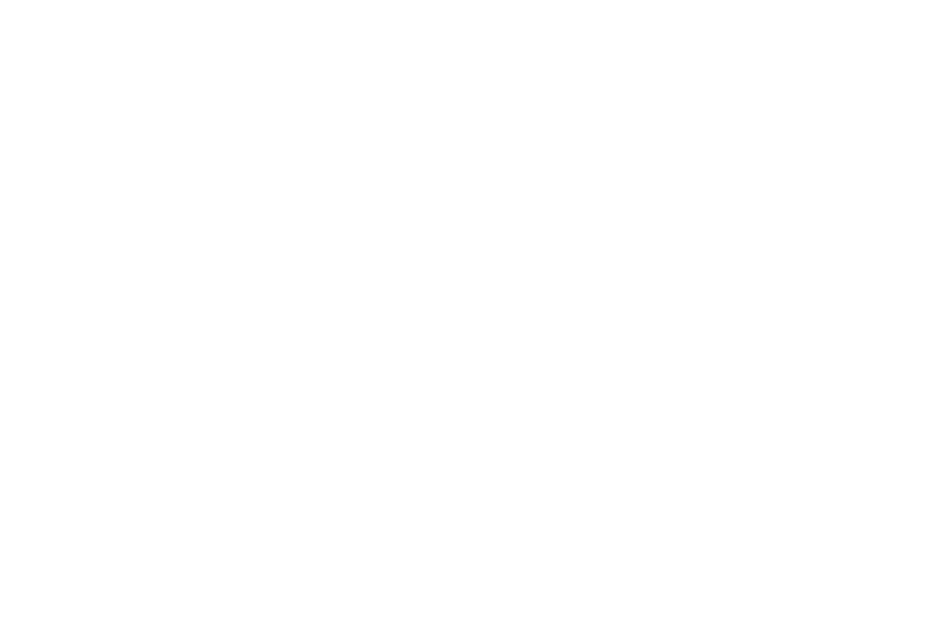}}\\[1ex]
\includegraphics[width=18.5cm]{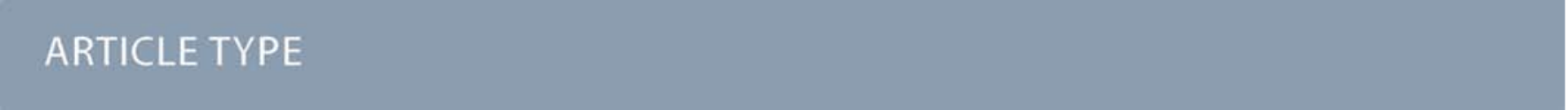}}\par
\vspace{1em}
\sffamily
\begin{tabular}{m{4.5cm} p{13.5cm} }

\includegraphics{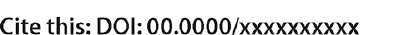} & \noindent\LARGE{\textbf{Unravelling the Full Relaxation Dynamics of Superexcited Helium Nanodroplets
}} \\
\vspace{0.3cm} & \vspace{0.3cm} \\

 & \noindent\large{Jakob D. Asmussen,\textit{$^{a}$} Rupert Michiels,\textit{$^{b}$} Katrin Dulitz,\textit{$^{b}$} Aaron Ngai,\textit{$^{b}$} Ulrich Bangert,\textit{$^{b}$} Manuel Barranco,\textit{$^{c,d}$} Marcel Binz,\textit{$^{b}$} Lukas Bruder,\textit{$^{b}$} Miltcho Danailov,\textit{$^{e}$} Michele Di Fraia,\textit{$^{e}$} Jussi Eloranta,\textit{$^{f}$} Raimund Feifel,\textit{$^{g}$} Luca Giannessi,\textit{$^{d}$} Marti Pi,\textit{$^{c,d}$} Oksana Plekan,\textit{$^{d}$} Kevin C. Prince,\textit{$^{d}$} Richard J. Squibb,\textit{$^{g}$} Daniel Uhl,\textit{$^{b}$} Andreas Wituschek,\textit{$^{b}$} Marco Zangrando,\textit{$^{d,h}$} Carlo Callegari,\textit{$^{d}$} Frank Stienkemeier,\textit{$^{b}$} and Marcel Mudrich\textit{$^{a}$}} \\

\includegraphics{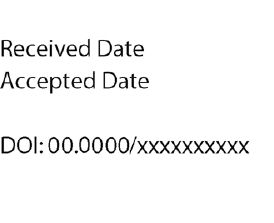} & \noindent\normalsize{The relaxation dynamics of superexcited superfluid He nanodroplets is thoroughly investigated by means of extreme-ultraviolet (XUV) femtosecond electron and ion spectroscopy complemented by time-dependent density functional theory (TDDFT). Three main paths leading to the emission of electrons and ions are identified: Droplet autoionization, pump-probe photoionization, and autoionization induced by re-excitation of droplets relaxing into levels below the droplet ionization threshold. The most abundant product of both droplet autoionization and photoionization is He$_2^+$, whereas the delayed appearance of He$^+$ is indicative of the ejection of excited He atoms from the droplets. The state-resolved time-dependent photoelectron spectra reveal that intermediate excited states of the droplets are populated in the course of the relaxation, terminating in the lowest-lying metastable singlet and triplet He atomic states. The slightly faster relaxation of the triplet state compared to the singlet state is in agreement with the simulation showing faster formation of a bubble around a He atom in the triplet state. 
} \\

\end{tabular}

 \end{@twocolumnfalse} \vspace{0.6cm}

  ]

\renewcommand*\rmdefault{bch}\normalfont\upshape
\rmfamily
\section*{}
\vspace{-1cm}


\footnotetext{\textit{$^{a}$~Department of Physics and Astronomy, Aarhus University, Denmark. Email: dall@phys.au.dk, mudrich@phys.au.dk}}
\footnotetext{\textit{$^{b}$~Institute of Physics, University of Freiburg, Germany }}
\footnotetext{\textit{$^{c}$~Departament FQA, Facultat de Física, Universitat de Barcelona, Spain }}
\footnotetext{\textit{$^{d}$~Institute of Nanoscience and Nanotechnology (IN2UB), Universitat de Barcelona, Spain }}
\footnotetext{\textit{$^{e}$~Elettra-Sincrotrone Trieste S.C.p.A., Italy }}
\footnotetext{\textit{$^{f}$~Department of
Chemistry and Biochemistry, California State University at Northridge, Northridge, CA 91330, USA }}
\footnotetext{\textit{$^{g}$~Department of Physics, University of Gothenburg, Sweden }}
\footnotetext{\textit{$^{h}$~CNR-IOM,~Elettra-Sincrotrone Trieste S.C.p.A., Italy }}




\section{Introduction}
Ultrafast energy relaxation of photoexcited nanosystems is a fundamental process occuring in the atmosphere~\cite{George2015}, in (radio)biology~\cite{Kleinermanns2013,Barbatti:2010,Son:2013}, and in photocatalytic materials~\cite{Papcke2020a}. Depending on the couplings of the internal degrees
of freedom, relaxation can be ultrafast, converting electronic excitations within a few femtoseconds (fs)~\cite{King2012} or slow, if the energy is trapped in a metastable state that decouples from its environment~\cite{McKinsey1999}. To unravel elementary steps in the relaxation path, a major challenge is the complexity of most condensed-phase systems. For this purpose, studying model systems with a simple structure and composition is paramount. For weakly-bound nanoparticles, superfluid helium (He) nanodroplets are ideal model systems due to their extremely weak interatomic binding, their homogeneous superfluid density distribution~\cite{Toennies2004a}, and the simple electronic structure of the He atom which greatly simplifies electron spectra. He nanodroplets are fascinating objects which have been extensively studied in terms of their quantum fluid properties~\cite{Grebenev:1998,Gomez:2014} and owing to their ability to pick up impurities in the form of atoms and molecules, for which they serve as a weakly perturbing cryomatrix~\cite{Stienkemeier2006}. Due to their very low temperature (0.37~K), He nanodroplets allow us to study the spectroscopy and chemistry of cold molecules and nanostructures~\cite{Mudrich:2004,MuellerPRL:2009,boatwright2013helium,mauracher2018cold,henning2019experimental}. Probing dynamical processes in superfluid He on the molecular scale has attracted particular interest, but has remained challenging both for experimentalists and theorists~\cite{Benderskii:2002,Ancilotto2017,Ziemkiewicz2015,Vangerow:2017,thaler2018femtosecond,chatterley2019long}.

The electronically excited states of He can be accessed through electron bombardment or extreme ultraviolet (XUV) radiation. Fluorescence spectroscopy using synchrotron radiation has revealed essentially two absorption bands below the atomic ionization threshold (24.6~eV)~\cite{Joppien1993}. The lower band centered at 21.6~eV consists of excited states related to the $1s2s$ and $1s2p$ atomic states of He, and the higher band centered at 23.8~eV can be associated mostly with the $1s3p$ and $1s4p$ atomic states. \textit{Ab intio} calculations and a model based on localized atomic Rydberg states support this interpretation of absorption spectra~\cite{Closser2010,Kornilov2011}. The ionization potential of He droplets falls between the two bands ($\sim23$~eV)~\cite{Frochtenicht1996,Peterka2003} and lies 1.6~eV below the atomic ionization potential. This autoionization channel opens due to interaction between a localized excited state (He$^*$) with a nearest neighbour atom coupling to the bound states of He$_2^+$ at short internuclear distance~\cite{Peterka:2007}.
Electrons created by this process were found to have nearly zero kinetic energy (ZEKE)~\cite{Peterka2003}. Dispersed fluorescence spectra recorded above the droplet autoionization threshold have shown that the formation of excited triplet states is not found below this threshold. Likely, these were formed by recombination of the electron with the He$_2^+$ produced by the droplet autoionization process~\cite{VonHaeften1997,Haeften:2005}. Direct information about triplet excitation spectra of He droplets can be obtained from energy-loss spectra of photoelectrons emitted by the droplets at photon energies $h\nu>44$~eV~\cite{Shcherbinin2019}. Previous XUV pump-probe time-resolved measurements~\cite{Ziemkiewicz2015,Mudrich} as well as \textit{ab initio} simulations~\cite{Closser2014} have revealed the following relaxation dynamics initiated by resonant He droplet excitation: The excited state localizes on an atomic (He$^*$) or molecular center (He$_N^*$,~$N=2,\,3$) within $\sim 100$~fs. Due to Pauli repulsion of the excited electron from the surrounding ground state He, a bubble forms around He$^*$ with a radius of 6.4~\AA~\cite{Hansen1972,VonHaeften2002}. 
If the excitation localizes close to the droplet surface, the bubble bursts before fully forming and ejects the He$^*$~\cite{Kornilov2011}. Otherwise, the excited state remains trapped inside the bubble in a metastable electronic state~\cite{VonHaeften2002}.

The goal of the present study is to extend our previous work~\cite{Mudrich} (where we excited He droplets to the lowest absorption band) to the excitation of the higher $1s3p/1s4p$-correlated absorption band lying above the droplet ionization threshold using tunable XUV pulses generated by a seeded free-electron laser (FEL). This study complements the work done by the group of O. Gessner and D. Neumark at Berkeley, where a similar pump-probe scheme was applied based on high-harmonic generation (HHG) of intense laser pulses~\cite{Ziemkiewicz2014}. In our study, we are now able to trace the full relaxation dynamics of the excited droplets on a time scale of 0.1~ps up to 150~ps owing to (i) more efficient probing of the dynamics using intense near-ultraviolet (UV) probe pulses and (ii) use of a spectrometer capable of simultaneously detecting high-resolution electron energy distributions and ion time-of-flight (TOF) spectra. We identify essentially three relaxation pathways, one of which has not been observed so far: The efficient population of the lowest triplet state of atomic He ($1s2s\,^3$S). The relaxation dynamics is modeled using a set of coupled rate equations, from which we infer the formation times of various intermediate states. The latter are compared to the results of TDDFT calculations for the lowest two excited states of atomic He ($1s2s\,^1$S/$^3$S).  

\section{Methods}

\subsection{Experimental setup}
\label{sec:Methods}
The experiments were performed at the low density matter (LDM) endstation of the seeded FEL FERMI in Trieste, Italy~\cite{Lyamayev2013}. The FEL was seeded with the third harmonic of a Ti:Sapphire laser (261~nm) providing photons with an energy of 23.7~eV (FEL undulators tuned to the 5$^{th}$ harmonic of the seed laser)~\cite{Allaria2012}, matching the center position of the $1s3p/1s4p$ absorption band of He droplets. The temporal duration of the FEL pulses was 80~fs (FWHM), and their intensity was varied using different filters ($10^9$-$10^{10}$~Wcm$^{-2}$). The beam was focused to a ($70~\mu$m)$^2$ spot size (FWHM) in the interaction region~\cite{raimondi2019kirkpatrick}. The UV probe laser pulses were generated from the second harmonic of the Ti:Sapphire seed laser (3.2~eV), and the pulse intensity was set to $3\times10^{11}$,  $1.3\times10^{12}$, and $3\times10^{12}$~Wcm$^{-2}$. The cross correlation of the pump and probe pulse was experimentally determined to be 240~fs FWHM. 

He nanodroplets were formed by expanding He gas from a high-pressure reservoir (50~bar) through a pulsed, cryogenically cooled Even-Lavie-type valve at a pulse repetition rate of 50~Hz~\cite{Pentlehner2009}. The mean size of the droplets was varied in the range of $10^4$-$10^5$ atoms per droplet by varying the valve temperature between 14 and 18~K. 
The droplet beam was crossed with the laser beam at the center of a combined magnetic-bottle-type electron spectrometer and ion TOF spectrometer used to record photoelectron spectra (PES) and ion mass-over-charge spectra simultaneously. Both the electron and the ion-TOF spectra were measured in a single-shot detection mode and individual ion hits were discriminated to eliminate electronic noise 

To obtain ion kinetic energy information from the TOF spectra, a matrix inversion method converting between these two quantities was used. With a pre-generated basis set of TOF functions, where each function corresponds to a specific ion kinetic energy, we construct a least-squares matrix inverse~\cite{penrose1956best}, regularized with a Tikhonov (or ridge regression) parameter. The basis TOF distribution functions were obtained through numerical simulations of ion trajectories within our known spectrometer geometry and its known electric/magnetic fields, performed using the program SIMION~\cite{dahl2000simion}. The starting points of the trajectories were found through a calibration procedure. This starting geometry is determined where the mismatch between simulated and observed gas-phase (zero initial velocity) He$^+$ spectra is minimized through fitting the following constraints: The width of the collimated droplet beam, the width of the laser beam perpendicular to the droplet beam, and their location within the spectrometer. These He$^+$ calibrated trajectories were then extrapolated to include initial velocity vectors in all directions, up to 2~eV, and for He$_2^+$ and He$_3^+$ ions. Finally, these trajectories were then combined to yield our TOF basis functions, uniquely determined by mass and charge. 

\subsection{Time-dependent density functional simulations}
The dynamics of the excited He droplet was simulated using TDDFT for a bulk superfluid density distribution~\cite{Ancilotto2017} to which the dynamics of an excited He atom (He$^*$) is self-consistently coupled. These calculations were done for He$^*$ in the lowest singlet and triplet excited states $1s2s\,^1S/^3S$. The dynamics of the He$^*$ ``impurity'' was followed by solving the Schr\"odinger equation for the He$^*$ wave function $\Psi(\mathbf{r},t)$, where the potential term is given by the He$^*$-liquid interaction, which is obtained by convoluting the liquid atom density $\rho(\mathbf{r},t)$ with the He$^*$-He pair potential $V_{\rm{He-He^*}}(r)$~\cite{fiedler2014interaction,Mudrich}. 
Details on how the coupled TDDFT-Schr\"odinger equations are solved can be found in Refs.~\cite{Ancilotto2017,laforge2020time}. The interaction energy $U^*(t)$ of He$^*$ with its superfluid local environment in the (time-dependent) initial state was computed as
 \begin{equation}
 U^*(t) = \int\int d\mathbf{r} d\mathbf{r'} \, |\Psi(\mathbf{r'},t)|^2  \, V_{\rm{He-He^*}}(|\mathbf{r'}-\mathbf{r}|) \, \rho(\mathbf{r},t) \; .
 \end{equation}
The interaction energy of He$^+$ with the superfluid final state, $U^+(t)$, is obtained in the same way using instead  $V_{\rm{He-He^+}}(|\mathbf{r'}-\mathbf{r}|)$.
The energy transferred to the photoelectron is obtained as
\begin{equation}
T_e(t) = 2h\nu' -[U^+(t)-U^*(t)] \; , 
\end{equation}
where we take twice the probe photon energy $h\nu'$ as it takes two photons to reach above the vertical ionization threshold of He. We calculated the average radius of the spherical bubble ($R_b$), which is formed around He$^*$ as~\cite{Ancilotto2017}
\begin{equation}
   R_b(t) = \left[ \frac{3}{4\pi}\int d\textbf{r} \;\left( 1 - \frac{\rho(\textbf{r},t)}{\rho_0} \right) \right],
 \end{equation}
where $\rho_0$ is the bulk superfluid He density.

\section{Results and discussion}
In this experiment, PES and ion TOF traces were measured simultaneously as a function of pump-probe delay for three different He droplet sizes and three different intensities of the probe pulses, see Sec.~\ref{sec:Methods}. In the first part of this paper, we discuss the measured yields of electrons and ions as they reveal the dominant relaxation processes and their dynamics. The time-resolved PES give detailed insight into the evolution of the internal energy of the system. The time-resolved ion mass spectra and kinetic energies provide complementary information about the final state of fragmentation of the excited He droplets. By comparing PES and ion spectra we obtain a complete picture of the relaxation dynamics.

\begin{figure}[t]
\centering
  \includegraphics[width = \columnwidth]{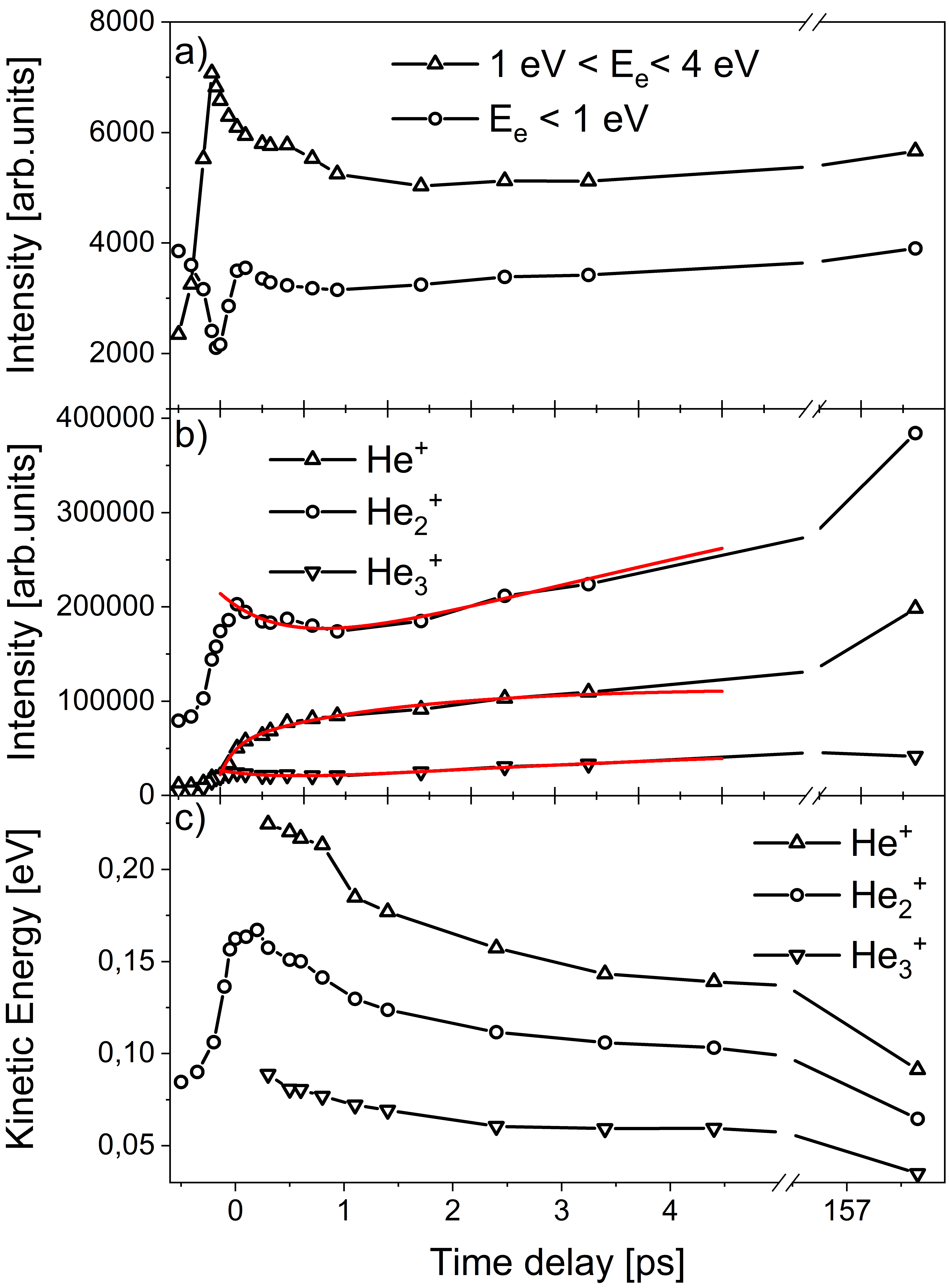}
  \caption{a) Electron yields in two different energy ranges; b) yields of the most abundant He ions. The thick red lines depict fits to the data (see text for details). c) Mean kinetic energies of He ions inferred from the peak profiles of the TOF spectra.}
  \label{ion_18uJ}
\end{figure}
\subsection{Yields of electrons and ions}
We start the discussion of the experimental data with the yields of electrons and ions, shown in Fig. \ref{ion_18uJ} a) and b) for droplets with an average size of $N_\mathrm{He}=41,000$. The FEL pump pulses and the UV probe pulses have intensities of $1\times10^{10}$~Wcm$^{-2}$ and $1.3\times10^{12}$~Wcm$^{-2}$, respectively. Two ranges of electron kinetic energy $E_e$ are considered separately: Electrons with low energy $E_e<1$~eV are mostly generated by autoionization of the droplets due to non-adiabatic coupling of the superexcited droplet state with the He$_2^+$ ionic state~\cite{Peterka2003,Peterka:2007}. Electrons in the energy range $1$~eV\,$<E_e<4$~eV are generated by pump-probe resonant two-photon or three-photon photoionization. The low-energy electrons (open circles) show a rather constant level of intensity for pump-probe delays varying from negative (UV first, XUV second) to positive values (XUV first, UV second). At negative delay, the He droplets are only resonantly excited by the XUV pulse into the $3p/4p$ absorption band of He droplets~\cite{Joppien1993}, as He droplets in their ground state are transparent to the UV pulse. In this autoionization regime, He$_2^+$ ions are the only detected fragments. Their kinetic energy [Fig.~\ref{ion_18uJ} c)] is low (0.08~eV) as they are formed in low-lying vibrational states. The vibrational energy is partly converted into kinetic energy of the ejected ion in the course of vibrational relaxation~\cite{callicoatt1998fragmentation}.


A narrow dip in the low-energy electron yield in the region of the temporally overlapping pump and probe pulses indicates the depletion of the population in the superexcited state above the autoionization threshold. This signal drop is compensated by a steep rise of the yield of photoelectrons ($1$~eV~$ < E_e < 4$~eV), which is followed by a decrease until about 2~ps delay and a subsequent slow increase. This trend is similar for the He$_2^+$ and He$_3^+$ ion yields with some time delay. Thus, He$_2^+$ are also the dominant ions ejected from He droplets in the regime of droplet photoionization, in accordance with previous static experiments~\cite{BuchtaJCP:2013}. The higher He$_2^+$ mean kinetic energy of up to 0.17~eV [Fig.~\ref{ion_18uJ} c)] matches previous measurements~\cite{shcherbinin2017interatomic,Shcherbinin2019} and reflects the higher vibrational excitation of He$_2^+$ formed by photoionization~\cite{Peterka:2007}. 

The He$^+$ atomic ions appear more slowly and their yield rises monotonically. The delayed rise of the yield as well as the higher He$^+$ kinetic energies of up to 0.23~eV point at a different ejection mechanism of He$^+$ than for He$_2^+$: Upon excitation of the He droplet, a He$^*$ excited atom is impulsively ejected from the droplet due to Pauli repulsion acting between the He$^*$ and the He droplet~\cite{BuenermannJCP:2012}. Note that the pump-probe dynamics of He$^+$ closely resembles the one measured for resonantly excited alkali atoms residing at the surface of He droplets~\cite{Vangerow:2015,Vangerow:2017}, and for excited indium atoms located inside the droplets~\cite{thaler2018femtosecond}. Those experiments have shown that pump-probe ion traces are determined by the competing dynamics of ejection of the excited atom and the ion falling back into the droplet when it is created at short delays. From the fit of the He$^+$ traces in our experiment, we obtain two rise times, $0.18$~ps and $2.0$~ps. We identify the shorter time with the detachment of He$^*$ located near the surface of the droplet, whereas the longer one is associated with He$^*$ ejected from deeper inside the droplet. These rise times are somewhat longer that those reported by the Berkeley group~\cite{Kornilov2011}. In those experiments, He$^*$ in Rydberg states were directly emitted from the droplet surface. Owing to the larger atomic orbitals of Rydberg atoms, repulsion from the droplet is stronger than for He$^*$ in $n=2$, and ejection is faster. In contrast, no ejected He atoms in Rydberg states are detected in our experiment (see the next section). A measurement at long pump-probe delay (158~ps) reveals even higher yields of He$^+$ and He$_2^+$ ions, as well as drastically reduced kinetic energies. This indicates that an additional slow mechanism is active, which leads to the detachment of slow excited He atoms, He$^*$, and excimers, He$_2^*$, by a process akin to thermal evaporation~\cite{Vangerow:2017}. As these species are photoionized after detaching from the droplets, the resulting photoions can no longer fall back into the droplets and thus elude detection. This explains the increased yield of ions versus electrons at long pump-probe delay. Note that we cannot rule out the formation of He$_2^*$ in the singlet groundstate as the photoelectron spectrum is indistinguishable from that of singlet He$^*$.

The time-resolved PES, discussed in the next section, show that the $1s3p/1s4p$-excited He droplets electronically relax into the lower $2s/2p$ droplet band by interband relaxation into low-lying states within about 200~fs, as previously observed~\cite{Ziemkiewicz2014}. Thus, the dip in the low-energy electron yield and the concurrent peak in the photoelectron yield reflects the time window in which resonant 1+1'-photon ionization is possible. At any longer delay, the yields of electrons and ions result from autoionization occurring within this short time window and from 1+2'-photon ionization. In addition, the droplets  that have relaxed into $n=2$ states are re-excited by the delayed probe pulse, causing them to autoionize~\cite{Kornilov2011}. The UV probe-photon energy (3.2~eV) matches well with the gap between the lower edge of the $2s/2p$ band at 21~eV and the maximum of the $3p/4p$ band around 24~eV. The delayed increase of He$_2^+$ and He$_3^+$ yields at delays $\gtrsim1$~ps, which roughly follows the yield of low-energy electrons, is attributed to this process. This interpretation is supported by the He$_2^+$ kinetic energy, which drops to nearly the value at negative delays where autoionization is the only possible decay path. Thus, the slight decrease of the He$_{2,\,3}^+$ signals at delays $\lesssim1$~ps is due to the relaxation of the excited droplets below the 1+1'-photon ionization threshold into states that are subsequently re-excited by resonant absorption of a UV-probe photon.

The thick red lines in Fig.~\ref{ion_18uJ} b) show the result of fitting three data sets simultaneously, each recorded at different probe pulse intensities. The fit model for the ion yield data is
\begin{align}
    I_{\text{He}^+}(t) &= c_1(1-e^{-t/\tau_1})+c_2(1-e^{-t/\tau_2}) \\
     I_{\text{He}_2^+}(t) &= c_3(1-e^{-t/\tau_3})+c_4e^{-t/\tau_4}\\
     I_{\text{He}_3^+}(t) &= c_5(1-e^{-t/\tau_3})+c_6(1-e^{-t/\tau_4})+c_7e^{-t/\tau_5},
\end{align}
where $I_{\text{He}_n^+}$ is the yield intensity for the He monomer, dimer and trimer cation. The initial drop of He$_2^+$ and He$_3^+$ yields proceeds with a time constant of $0.85$~ps, whereas for the delayed signal rise we find that the dimer yield proceeds with a time constant of $20$~ps, and that the triplet yield rise can be described with time constants $2.0$ and $20$~ps.

\begin{figure*}[h]
\centering
  \includegraphics[width = 2\columnwidth]{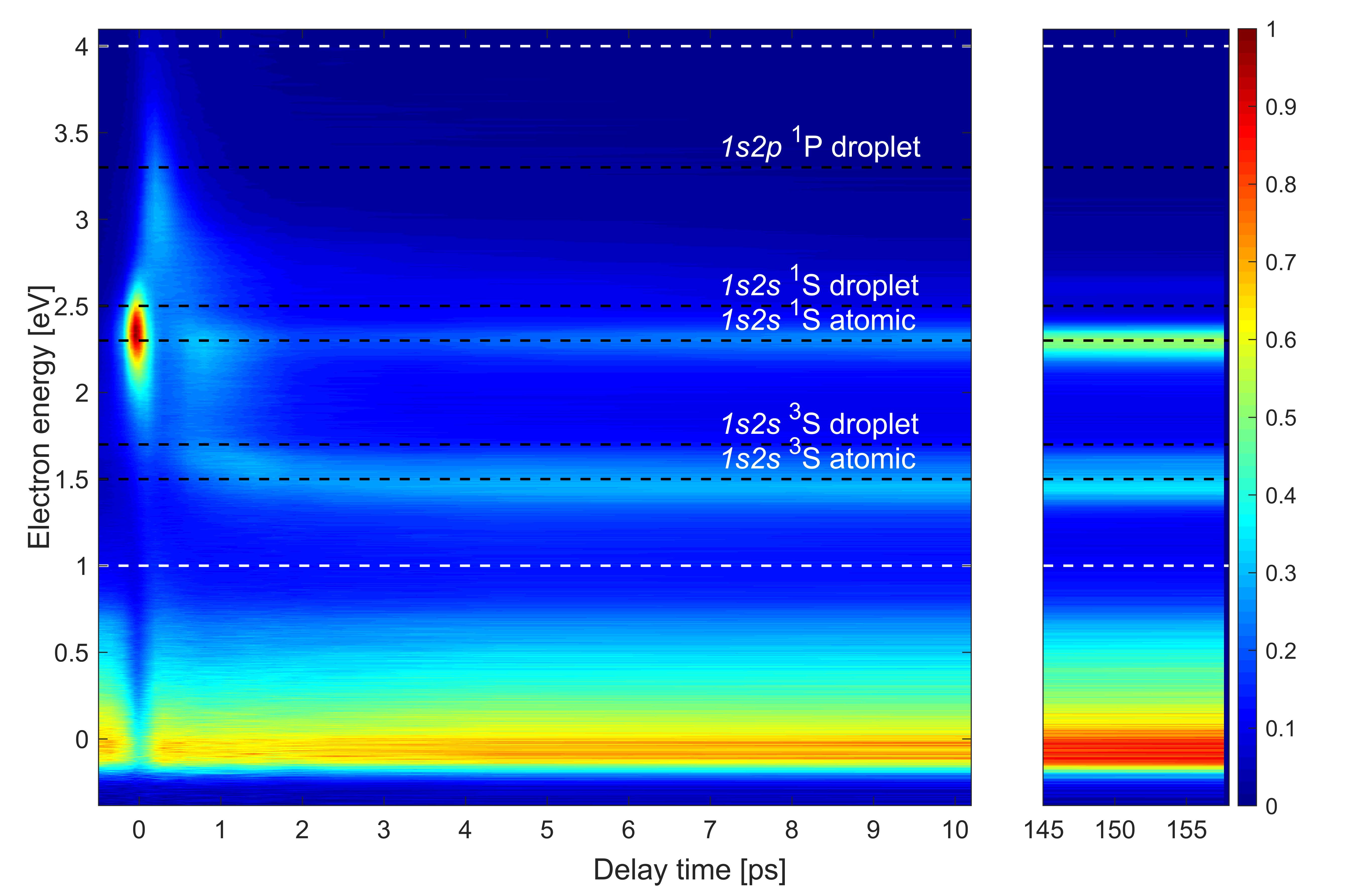}
  \caption{Time-resolved photoelectron spectra for He droplets containing on average $41,000$ atoms. The density plot represents 22 spectra at different pump-probe delays. The white lines indicate the region fitted with a multi-Lorentzian function. The horizontal black lines indicate expected positions of atomic and droplet~\cite{Shcherbinin2019} states of He for two-photon ionization. The intense feature around zero delay is due to one-photon ionization out of the $1s3p/1s4p$ droplet state and therefore does not correspond to any of the black lines. }
  \label{2D_data}
\end{figure*}
\subsection{Time-resolved electron spectra}
To obtain more detailed insight into the relaxation dynamics of the excited states of the He droplets, we now turn to the time-resolved PES. Figure~\ref{2D_data} shows a compilation of 22 PES recorded as a function of pump-probe delay for He droplets of mean size $N_\mathrm{He}=41,000$, pump pulse intensity $1\times10^{10}$~Wcm$^{-2}$ and probe pulse intensity $1.3\times10^{12}$~Wcm$^{-2}$. Clearly, a large fraction of electrons are emitted at very low energy $< 0.5~$eV due to droplet autoionization. Only in the delay range of overlapping pump-probe pulses is this ZEKE signal notably depleted.

At higher electron energies $E_e > 0.5~$eV, the PES reveal the population of different discrete electronic states. Part of the population in the upper droplet band ($3p/4p$) is directly photoionized by one UV photon (bright spot at $E_e = 2.3$~eV near zero delay) while the lower band ($1s2s/1s2p$) and the atomic states $1s2s\,^1S$ (2.4~eV) and $1s2s\,^3S$ (1.6~eV), populated by electronic relaxation within $\lesssim 1~$ps, are ionized by two-photon absorption. Similar time-resolved PES were previously measured by the Berkeley group in an XUV/UV pump-probe experiment using HHG to generate the pump pulses~\cite{Ziemkiewicz2014}. Although the spectra also contained two faint bands at electron energies corresponding to two-photon ionization of the $1s2s\,^{1,\,3}S$ states, the authors interpreted them differently. Presumably due to the rather low probe pulse intensity they used, they did not take two-photon ionization into consideration. Furthermore, due to the lower resolution of the electron spectrometer that was used, the bands were spectrally broadened and the correct assignment of the two bands to the $1s2s\,^{1,\,3}S$ states was not as compelling as in the present case.

Similarly to our previous XUV/UV pump-probe study~\cite{Mudrich}, we find that the relaxation dynamics is entirely determined by droplet-induced electronic relaxation into the lowest excited states of the He atom, $1s2s\,^{1}S$ and $^{3}S$. No ejected Rydberg atoms are detected, in contrast to previous observations when using a near-infrared (NIR) pulse as a probe~\cite{Kornilov2011}. The different findings for NIR and UV probing are likely related to the $\sim1.1$~eV potential energy barrier that electrons have to overcome in order to leave the He droplet bulk~\cite{Ziemkiewicz2014,Broomall1976,Wang}. Thus, NIR photoionization effectively singles out excited atoms in Rydberg states ($1s4p$, $1s3d$) that are ejected from the droplets, as they can be ionized by one NIR photon, whereas detection of lower-lying levels is suppressed.

\begin{figure}[h]
\centering
 \includegraphics[width = \columnwidth]{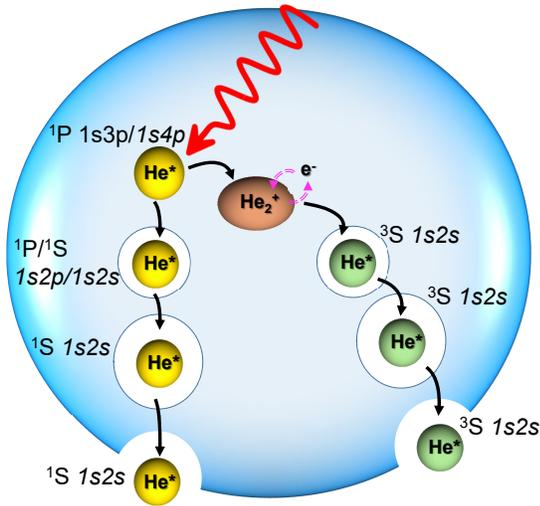}
  \caption{Illustration of the proposed relaxation dynamics after excitation of He droplets into the $1s3p/1s4p$ band. The dynamics splits into two paths leading to the $1s2s\,^1S$ and $^3S$ metastable atomic states. The $^1S$ singlet path follows intra- and interband relaxations through electronic droplet states whereas the $^3S$ triplet state is formed by recombination after autoionization. A bubble forms around the excited state which bursts when the excited state moves to the surface and releases the metastable atom. }
  \label{cartoon}
\end{figure}
The high temporal and spectral resolution achieved in this work allows us to attempt to model the full time evolution of excited He droplets by a closed rate-equation model. Figure~\ref{cartoon} illustrates the various relaxation steps taken into account in our model. Essentially, the relaxation consists of two different relaxation paths leading to the final $1s2s\,^{1,3}S$ metastable states. The path to the $1s2s\,^1S$ state is a step-wise process through an intermediate excited droplet state. Most likely, the excited atom undergoes a fast intraband relaxation to the bottom of the upper droplet band before interband relaxation to the lower band occurs. Intraband relaxation in the upper band happens faster than what we can resolve in the experiment. In the lower droplet band ($1s2s/p$), we see intraband relaxation to the $1s2s$ droplet state while a bubble forms around the excited state. Thereafter, the bubble moves to the surface of the droplet, burst and thereby releases the excited atom which either leaves the droplet or remains weakly attached to the droplet surface. Here, we assume that the relaxation dynamics of the lower band is similar to the one described in our previous FEL study, where we directly excited into the lower band~\cite{Mudrich}.

For relaxation into the metastable $1s2s\,^3S$ state, our model follows the discussion in the literature~\cite{VonHaeften1997}; triplet states are formed by recombination of electrons and ions after droplet autoionization. The time-resolved PES reveal a time gap between depletion of the upper droplet band ($3p/4p$) and the appearance of the droplet triplet states. This indicates an intermediate step between the two states, which matches well with the description of relaxation through autoionization followed by recombination. Presumably, He$_2^+$ is the precursor cation and dissociative recombination (DR) of He$_2^+$ leads to one atom in the He ground state ($1s^2\,^1S$) and the other atom in a metastable state~\cite{Carata1999,Royal2005a}. A multichannel quantum defect theory (MQDT) calculation showed that for He$_2^+$ in the vibrational ground state, DR from ion-electron collisions with low-energy electrons ($>0.1$~eV) favors the population of dissociative states with $^3\Sigma_g$ symmetry over states with $^1\Sigma_g$ symmetry by two orders of magnitude~\cite{Carata1999}.  However, an experimental study showed branching ratios of $\sim60$\,\% and $\sim38$\,\% for electron capture into channels where the $1s2p\,^3P$ and the $1s2s\,^1S$ atomic states are formed, respectively. Only a small fraction of atoms are formed in the $1s2s\,^3S$ states, which would be expected from theory~\cite{Pedersen2005}. This was explained by non-adiabatic couplings which the theory does not include~\cite{Pedersen2005}. Since we do not detect any population in higher-lying triplet states than the $1s2s\,^3S$ state, this suggests that such non-adiabatic couplings are suppressed in the droplet environment. For this reason and to keep our model simple and robust, we do not include the formation of excited singlet states from autoionization and electron-ion recombination in our relaxation model. In the gas phase, non-dissociative recombination is a weak channel, but in the liquid He environment, the formation of He$_2^*$ excimers by recombination is well established~\cite{Benderskii:2002}. However, ionization of the triplet ground state of He$_2^*$ ($^3\Sigma_u^+$) would result in a photoelectron energy of $\sim2$~eV~\cite{Sunil1983} which is not present in the PES.

Whether autoionization followed by recombination is the only relaxation channel to form triplet states, is uncertain. We have recently shown that Penning ionization of doped alkali metal atoms on He droplets occurs from both the He $1s2s\,^1S$ state and, to a small extent, the $^3S$ state, after excitation into the lower absorption band of the droplet (21.6~eV)~\cite{BenLtaief2019}. This suggests the existence of a relaxation channel to the  $1s2s\,^3S$ state from states below the droplet ionization potential as well. We mention that spin quenching was previously observed for barium atoms attached to the surface of argon clusters~\cite{awali2016multipronged}. However, adding such a relaxation channel to the rate model does not improve the correlation between model and data. Therefore, it is omitted from our fit model.
\\
\\
\begin{figure}[h]
\centering
  \includegraphics[width = \columnwidth]{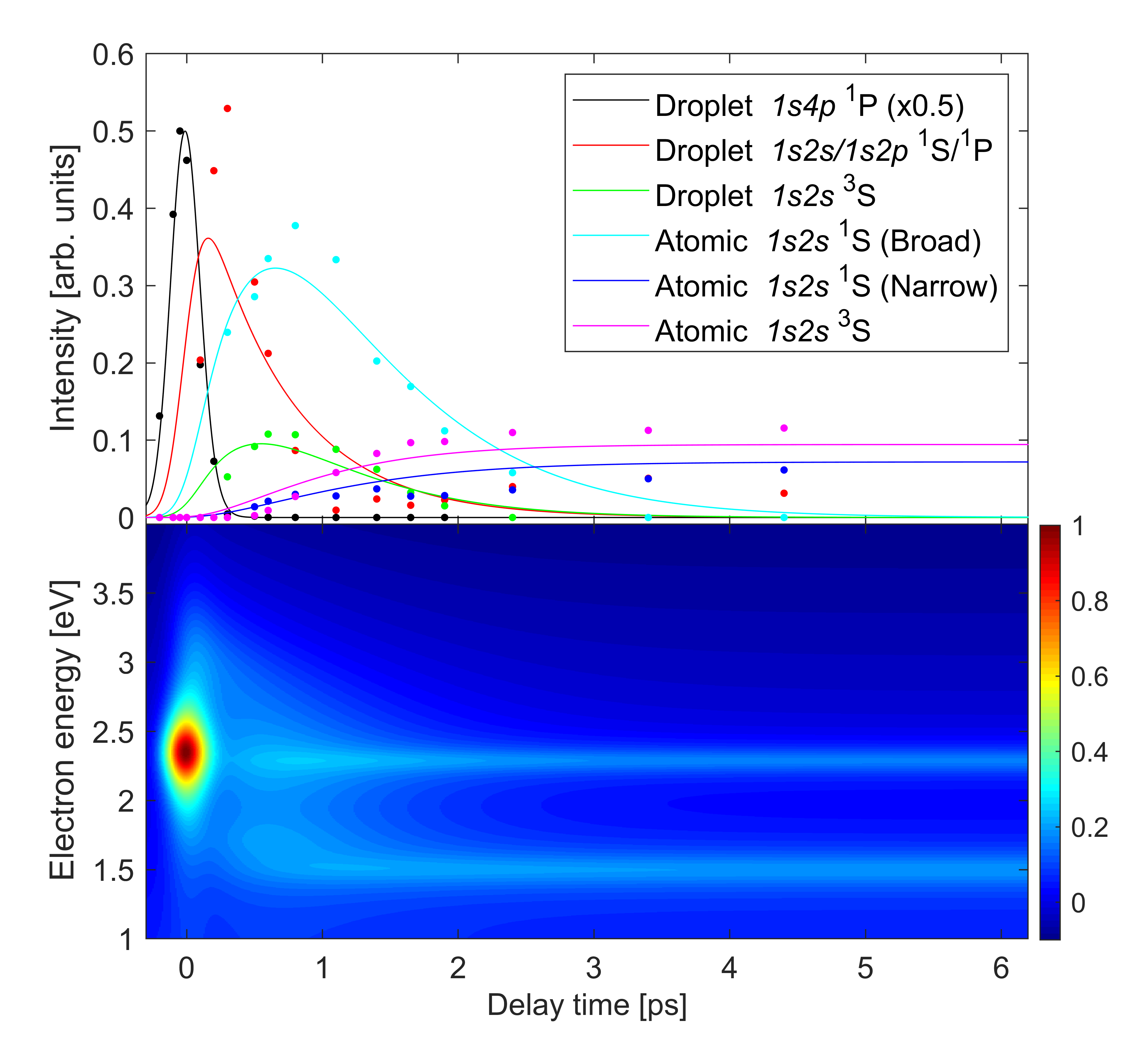}
  \caption{Results obtained from fitting a rate-equation model to the electron spectra. (top) The points show the peak intensity from a multi-Lorentzian fit to the experimental data for different delay times. The rate equation model is fitted to the intensities and the results of these fits are shown as lines. (bottom) A reconstruction of time-resolved electron spectra based on the fits. }
  \label{fit}
\end{figure}
For a quantitative analysis of the relaxation dynamics, the electron spectra are fitted with a multi-Lorentzian function consisting of 6 peaks. We fit this function in the $E_e$ interval between 1 and 4~eV (indicated by the white lines in Fig. \ref{2D_data}). The 6 peaks are ascribed to the three droplet states $1s3p/1s4p\,^1P$, $1s2s\,^1S / 1s2p\,^1P$, and $1s2s\,^3S$. We ascribe the other three peaks to the $1s2s$ singlet and triplet states, where two peaks correspond to the $1s2s\,^1S$ state representing a broad feature present at $\sim1$~ps pump-probe delay and a narrow feature at longer delays. We assign the broader feature to an intermediate state where the atomic state is perturbed by the droplet environment. Each peak is centered at the literature value of $E_e$~\cite{Shcherbinin2019}, except the $1s2s\,^1S / 1s2p\,^1P$ peak which shifts in energy from 3.0~eV to 2.6~eV according to an exponential function of time to model the observed intraband relaxation. Appropriate peak widths are determined to match the width of the features in Fig.~\ref{2D_data} and kept constant for all pump-probe delays. 

The peak amplitudes resulting from the multi-Lorentzian fits are then fitted by a rate equation model to determine the lifetimes of intermediate states. The following set of rate equations is used to model the relaxation dynamics:
\begin{align}
dN_{1s3p/1s4p \; ^1P}^{(D)}/dt &= k_1 I - k_2 N_{1s3p/1s4p \; ^1P}^{(D)} - k_3 N_{1s3p/1s4p \; ^1P}^{(D)} \\
dN_{\text{He}_2^+}^{(D)}/dt &= k_3 N_{1s3p/1s4p \; ^1P}^{(D)} - k_4 N_{\text{He}_2^+}^{(D)} \\
dN_{1s2s/1s2p \; ^1S/^1P}^{(D)}/dt &= k_2 N_{1s3p/1s4p \; ^1P}^{(D)} - k_5 N_{1s2s/1s2p \; ^1S/^1P}^{(D)} \\
dN_{1s2s \; ^3S}^{(D)}/dt &= k_4 N_{\text{He}_2^+}^{(D)} - k_6 N_{1s2s \; ^3S}^{(D)} \\
dN_{1s2s \; ^1S}^{(A^*)}/dt &= k_5 N_{1s2s/1s2p \; ^1S/^1P}^{(D)} - k_7 N_{1s2s \; ^1S}^{(A^*)} \\
dN_{1s2s \; ^1S}^{(A)}/dt &= k_7 N_{1s2s \; ^1S}^{(A*)} \\
dN_{1s2s \; ^3S}^{(A)}/dt &= k_6 N_{1s2s \; ^3S}^{(D)}
 \end{align}
Here, the subscript to the state population, $N$, indicates the state, and the superscript indicates whether it is an excited state of the droplet ($D$) or the atom ($A$). The symbol $^*$ marks the broad intermediate state associated with the final $1s2s\,^1S$ atomic state. $I$ is the intensity of the exciting laser pulse modelled as a Gaussian function in time. The full width at half maximum (FWHM) of this function is taken as the measured cross-correlation profile of pump and probe pulses to account for the temporal broadening due to the width of the probe pulse without increasing the number of fit parameters. The triplet states are expected to form only through autoionization followed by recombination, leading to an intermediate state ($N_{\text{He}_2^+}^{(D)}$) which does not contribute to the PES in the 1-4~eV range of $E_e$. The rate of autoionization is determined from the rising edge of the ZEKE electron signal, reflecting the recovery of the autoionization channel following the depletion of the droplet excitation by the probe pulse at $\sim 0$~ps delay. We obtain a characteristic lifetime with respect to autoionization of $0.30 \pm 0.21$~ps. The previous HHG experiment reports a rise time of $0.14 \pm 0.04$~ps for the fast recovery of the ZEKE signal~\cite{Kornilov2010,Ziemkiewicz2015}. The larger uncertainty on the autoionization time reported here is due to an additional contribution to the ZEKE signal from one-photon ionization out of the upper part ($>21.6$ eV) of the $1s2s/1s2p$ band, which is not described by the exponential rise. Similarly to the previous finding, we observe a recovery of the ZEKE signal and a rise at longer delay times (Fig.~\ref{2D_data}). This is ascribed to re-excitation of population already relaxed into $n=2$ states back into states lying above the autoionization threshold by the probe pulse~\cite{Kornilov2010,Kornilov2011}.

Fig.~\ref{fit} shows the result of fitting the electron spectra in Fig.~\ref{2D_data}. Besides the different rate constants, the fit model also contains a weight factor for each state as a free parameter. The weight factor is relative to the one-photon ionization out of the $1s3p/1s4p$ band and accounts for the varying ionization cross-sections for the different states. The two-photon ionization in the droplet environment may be affected by transient shifting in or out of resonance of intermediate states. Therefore, we refrain from trying to infer populations of individual states from our fit. The weight factor for the droplet singlet and triplet states are similar ($\sim 0.4$) as is the weight factor for the atomic singlet and triplet states ($\sim 0.1$). The ionization cross-section of metastable helium is similar for both singlet and triplet states,~\cite{Stebbings1973} which matches well with our findings of similar weight factors for the two spin states. Furthermore, it is expected that ionization is more efficient in the droplet environment due to a broadening of the states, which explains the higher weight factor for droplet states.  
\begin{table}[h]
\centering
\begin{tabular}{r|c}
\multicolumn{1}{l|}{Transition} & $\tau$ [ps] \\ \hline
$3p/4p \; ^1P^{(D)} \xrightarrow{} 2s/2p \; ^1S^{(D)}/^1P^{(D)}$ & $0.19 \pm 0.06$ \\
$3p/4p \; ^1P^{(D)} \xrightarrow{} \text{He}_2^+{}^{(D)}$ & $0.30 \pm 0.21$ \\
$\text{He}_2^+{}^{(D)} \xrightarrow{} 2s \; ^3S^{(D)}$ & $0.60 \pm 0.26$ \\
$2s/2p \; ^1S^{(D)}/^1P^{(D)} \xrightarrow{} 2s \; ^1S^{(A^*)}$ & $0.34 \pm 0.05$ \\
$ 2s \; ^3S^{(D)} \xrightarrow{}  2s \; ^3S^{(A)}$ & $0.59 \pm 0.21$ \\
$ 2s \; ^1S^{(A)} \xrightarrow{}  2s \; ^1S^{(A^*)}$ & $2.01 \pm 0.37$
\end{tabular}
\caption{Time constants obtained by fitting the rate equation model to results of a separate multi-Lorentzian fit of the electron spectra. Displayed values are the mean values and the standard deviations of 6 time-delay scans, all recorded at the same probe laser pulse parameters.}
\label{lifetimes}
\end{table}
\\
\\
Table~\ref{lifetimes} summarizes the average characteristic decay times for the different transitions in the rate model for a probe pulse intensity of $3\times10^{11}$~Wcm$^{-2}$. The mean and standard deviation is determined from 6 time-delay scans of droplets with sizes in the range $10^4$-$10^5$ atoms and pump intensity in the range $10^9$-$10^{10}$~Wcm$^{-2}$. No significant trend from the varying droplet size or pump intensity is seen between the different scans. This indicates that the observed dynamics is rather generic and nearly independent of the droplet size and laser pulse parameters. Similar to what was seen in our previous study~\cite{Mudrich}, the dynamics appears to be mostly driven by ultrafast electronic relaxation and bubble formation occurring in the bulk of the He droplets.

The dynamics is similar for a larger probe pulse intensity of $1\times10^{12}$~Wcm$^{-2}$, however the ratio of signals from the atomic singlet to the atomic triplet state is 0.7 for the smaller probe pulse intensity and 1.8 for the larger probe pulse intensity. This is due to the fact that re-excitation is a one-photon process whereas ionization requires two photons. In the case of low probe intensity, re-excitation into states above the autoionization threshold becomes more dominant compared to photoionization, which thereby leads to more autoionization followed by recombination into triplet states.

The fastest relaxation step in the rate model is the interband relaxation from the upper electronic band to the lower electronic band ($3p/4p \; ^1P^{(D)} \xrightarrow{} 2s\,^1S^{(D)}/2p\,^1P^{(D)}$). From the XUV/UV pump-probe experiment using HHG, the relaxation out of the initial highly excited state was determined to occur on a timescale of $\sim100$~fs~\cite{Ziemkiewicz2014}. This is faster than what we are able to resolve in our experiment which explains the larger lifetime from our fit model. In the HHG experiment, the authors identified a second broad feature relaxing on a timescale of $\sim450$~fs~\cite{Ziemkiewicz2014}. From their time-resolved PES, it is seen that this feature can be compared with the droplet-to-atomic relaxation for the singlet state ($2s\,^1S^{(D)}/2p\,^1P^{(D)} \xrightarrow{} 2s \; ^1S^{(A^*)}$) for which we find a time constant ($\sim340$~fs) that is in good agreement with the reported one ($\sim450$~fs). We find that the droplet-to-atomic state relaxation happens on a similar timescale for the $^1S$ and $^3S$ state, which suggest a similar relaxation mechanism for the two. 

Furthermore, the relaxation dynamics in He droplets can be compared to the electronic relaxation dynamics of excited xenon clusters\cite{Serdobintsev2018}. In small xenon clusters ($10<N<150$), intraband relaxation was shown to happen at a $\sim0.075$~eV/ps rate which is significantly slower than the intraband relaxation rate we find in the lower electronic band of He droplets ($\sim1$~eV/ps). This is mostly due to the much faster motion of He atoms owing to their much smaller mass, as intraband relaxation is mostly determined by the bubble formation dynamics around the excitation. 

\begin{figure}[h]
\centering
  \includegraphics[width = 1.05\columnwidth]{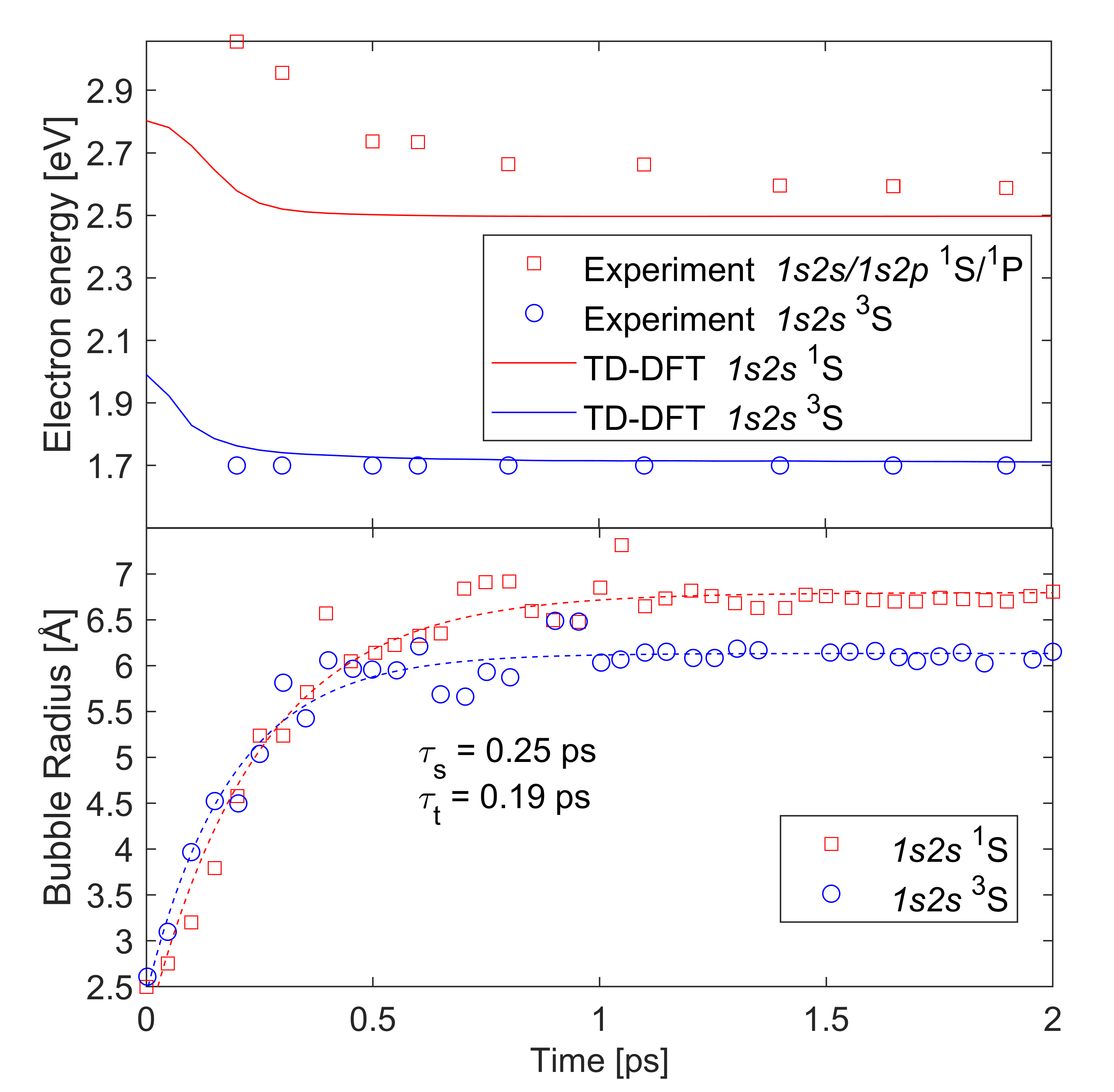}
  \caption{Results of TDDFT calculations. \textbf{(a)} The calculated kinetic energy of the photoelectrons decreases as the bubble forms around the metastable state and the bubble radius increases, shown in \textbf{(b)}. The calculated electron energies (solid lines in (a)) are compared to the fit values of the experimental spectra (symbols). The formation times of the bubbles is determined from exponential fits of the calculated bubble radius (dashed lines in (b)).}
  \label{TDDFT}
\end{figure}
The conspicuous new finding in this study is that both singlet and triplet metastable atomic states are populated by droplet relaxation, with slightly differing time constants. To get a better understanding of this final relaxation step, we have simulated the dynamics of these excited states in bulk superfluid He by means of TDDFT, as in our previous works~\cite{Mudrich,laforge2020time}. 
Following the excitation of a $1s2s\,^1S$ or $^3S$ state at time zero, a bubble forms around the excited state within about 300~fs~\cite{Mudrich}. Fig.~\ref{TDDFT} shows the results of these calculations. The formation times of the bubble are inferred from an exponential fit to the bubble radius. For the $^1S$ state, it is slightly longer (0.25~ps) than for the $^3S$ state (0.19~ps). This is likely due to the slightly larger asymptotic value of bubble size for the $^1S$ state (6.8~\AA) than for the $^3S$ state (6.1~\AA). The size of the bubble around the atom in the $^3S$ state is in good agreement with the 6.4~\AA~radius previously reported~\cite{Hansen1972}.  
From the TDDFT calculations we infer the time-dependent shifting of excited state energies and of ionization energies as detailed in Sec.~\ref{sec:Methods}. The latter shift upwards in energy is as previously found from model calculations of the formation of a bubble around an excited atom in bulk liquid He~\cite{Eloranta2002}. The resulting kinetic energy of photoelectrons can be compared to the fitted electron peak position for the droplet states in the experimental data. For the $^3S$ droplet state, the fitted peak is constrained to the expected energy value and does not vary with the delay. This energy is in good agreement with the asymptotic value of the TDDFT calculation. In the multi-Lorentzian fit, one peak is assigned to the $n=2$, $^1S$/$^1P$ singlet state of the droplet. The peak position follows an exponential decrease assumed in the fit model. The TDDFT calculation only includes excitation into the $1s2s\,^1S$ state and therefore does not describe electronic relaxation within the $^1S$/$^1P$ droplet band. The fast drop of the theoretical electron energy within 0.3~ps is solely due to the formation of a bubble, whereas the experimental value shifts due to intraband relaxation including both the transition from the $^1P$ to the $^1S$ state and bubble formation. 
Note that the bubble expansion might be delayed while the initially delocalized droplet excitation localizes on one He$^*$ atom within a few 100~fs~\cite{Closser2014}. On the other side, the transient population of a localized highly excited state ($1s3p$, $1s2p$) might accelerate the bubble expansion due to the stronger He$^*$-He repulsion compared to the $1s2s$ states. More sophisticated model calculations are needed to shed light on the coupled electronic-nuclear dynamics in excited He droplets. The fact that the experimental electron energy approaches the theoretical value only at delays $\gtrsim 1~$ps seems to indicate that bubble formation as well as the migration of the bubble to the droplet surface, which is a slower process and not included in the TDDFT calculations~\cite{Mudrich}, both determine the dynamics of photoelectron energy. 


The concept of a transient redistribution of populations, \textit{i.e.} droplet-bound He$^*$ (generating He$_{2,\,3}^+$) vs. free He$^*$ (generating He$^+$), as well as He$^*$ in state $n=2$ (seen as discrete photolines in PES) vs. $n=3,\,4$ (ZEKE signal) is supported by the total electron counts (sum of the two curves in Fig.~\ref{ion_18uJ} a)) remaining nearly constant for delays $>0.2$~ps. The slightly enhanced electron counts at 0.2~ps and at long delays are likely due to efficient one-photon processes: At short delays, the unrelaxed $3p/4p$ droplet state is directly ionized by one-photon absorption; at long delays, the population which relaxes into $n=2$ decouples from the droplet by forming a bubble. Consequently, spectral lines become narrow and one-photon re-excitation from the most abundant $1s2s\,^3S$ state (19.8~eV) into the $1s3p\,^3P$ state (23.0~eV) becomes fully resonant. The fact that resonant re-excitation enhances the ZEKE electron counts indicates that part of the metastable He$^*$ atoms actually remain attached to the He droplets, probably in dimples at the surface similar to alkali atoms. At a much later stage, these He$^*$ will likely form excimers and radiatively decay to the groundstate~\cite{Buchenau:1991}.

\begin{figure}[h]
\centering
  \includegraphics[width = 1.05\columnwidth]{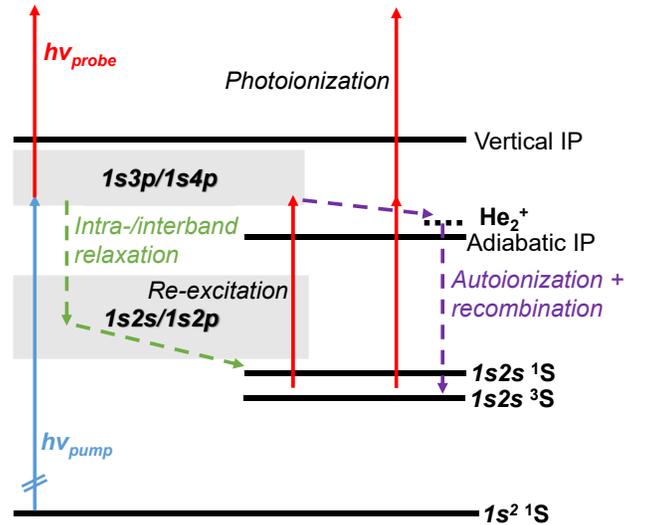}
  \caption{Schematic energy level diagram illustrating the pump-probe dynamics in He nanodroplets. The grey shaded areas represent the droplet bands, black horizontal lines are He atomic levels. Blue and red vertical arrows depict transitions driven by the pump and probe laser pulses, respectively. Green and purple dashed arrows indicate intraband/interband relaxation and autoionization, respectively. }
  \label{diagram}
\end{figure}

\section{Conclusion}
In conclusion, we have fully unravelled the relaxation dynamics of He droplets resonantly excited into the second, $3p/4p$-correlated absorption band by XUV pulses. We have identified essentially three ultrafast relaxation paths: (i) Direct autoionization generating He$_{2,\,3}^+$ and ZEKE electrons, (ii) autoionization followed by electron-ion recombination and relaxation to the metastable $1s2s\,^3S$ atomic state, and (iii) interband relaxation into the metastable $1s2s\,^1S$ atomic state. Fig.~\ref{diagram} schematically illustrates these processes in an energy-level diagram. 
The previously reported ejection of He Rydberg atoms was not observed and therefore seems to be a minor channel. From the fit of a rate equation model to the spectra, we obtain time constants for the individual relaxation steps, ranging between about 0.2~ps and 2~ps. In the final relaxation steps, the nearly unperturbed atomic $1s2s\,^{3}S$ state is formed slightly faster than the $1s2s\,^{1}S$ state, which may be related to the faster formation of a void bubble around these excited states predicted by TDDFT simulations in bulk superfluid He.
These dynamics were found to be very robust with respect to variations of the experimental conditions such as He droplet size and laser-pulse parameters.

While electron spectra directly reveal the evolution of the energy structure of the system, ion mass spectra inform about the final state of relaxation at a given pump-probe delay. The dominant ion, He$_{2}^+$, is predominantly formed by autoionization of the XUV-excited droplets and is enhanced through time-delayed re-excitation by the UV probe pulse. To a lesser extent, He$_{2}^+$ and He$_{3}^+$ is formed by direct pump-probe photoionization. The monotonously rising count rate of He$^+$ atomic ions indicates the ejection of free $n=2$-excited He atoms both from the droplet surface and from deeper within the bulk.

It would be very interesting to apply the same experimental scheme to the lowest, $2s/2p$-correlated absorption band. In our recent experiment~\cite{Mudrich}, the XUV-pump pulse was tuned to that band, but a weaker UV pulse was used that was unable to ionize the $1s2s\,^3S$ state by one-photon absorption. Thus, it remains an open question whether and to what extent the droplet environment induces spin quenching in the $n=2$ state of He. Other rare gas clusters should be studied to better understand their relaxation dynamics as well. There, surface and bulk excited states can be addressed selectively in valence-shell transitions~\cite{wormer1996evolution} and even by exciting the inner-shell electrons with X-ray pulses~\cite{bjorneholm1995core}. In this way, aspects of the relaxation dynamics that are specific to the quantum fluid properties of He droplets could be identified. Furthermore, multiply-excited and doped clusters/nanodroplets present additional rich phenomenology~\cite{laforge2019highly,laforge2020time,awali2016multipronged} to be explored using the new XUV and X-ray sources that are nowadays available.


\section*{Conflicts of interest}
There are no conflicts to declare.

\section*{Acknowledgements}
The authors are grateful for financial support from the Deutsche Forschungsgemeinschaft (DFG) within the project MU 2347/12-1 and STI 125/22-2 in the frame of the Priority Programme 1840 QUTIF, and from the Carlsberg Foundation. R.F. thanks the Swedish Research Council (VR) and the Knut and Alice Wallenberg Foundation for financial support.



\balance


\bibliography{He_Relaxation_PCCP} 
\bibliographystyle{rsc} 

\end{document}